\documentclass{svjour2}                    
\smartqed  
\usepackage{graphicx}
 \journalname{General Relativity and Gravitation}
\begin{document}

\title{Viscous Dark Energy Models with Variable $G$ and $\Lambda$}
 \subtitle{}


\author{Arbab I. Arbab}


\institute{Department of Physics, Faculty of Science, University
of Khartoum, P.O. Box 321, Khartoum 11115, Sudan
\\
 Department of Physics and Applied
Mathematics, Faculty of Applied Sciences and Computer, Omdurman
Ahlia University, P.O. Box 786, Omdurman,
Sudan\\\email{aiarbab@uofk.edu}}

\date{Received: date / Accepted: date}
\maketitle
\begin{abstract}
We consider a cosmological model with bulk viscosity ($\eta$) and
variable cosmological  $(\Lambda\propto \rho^{-\alpha},\ \
\alpha=\rm const.$) and gravitational ($G$) constants. The model
exhibits many interesting cosmological features. Inflation
proceeds du to the presence of bulk viscosity and dark energy
without requiring the equation of state $p=-\rho$. During the
inflationary era the energy density ($\rho$) does not remain
constant, as in the de-Sitter type. Moreover, the cosmological and
gravitational constants increase exponentially with time, whereas
the energy density and viscosity decrease exponentially  with
time. The rate of mass creation during inflation is found to be
very huge suggesting that all matter in the universe was created
during inflation. 
\end{abstract}
 \maketitle
\section{Introduction}
The present acceleration  of the universe as favored by the
supernovae data can be explained by hypothizing some exotic matter
dominated the present  universe evolution that breaks the strong
energy condition, viz., $p+3\rho\ge0$ (Reiss \emph{et al.},
Perlmutter \emph{et al.}). One variant of this exotic matter is
the one violates the null energy condition, viz., $p+\rho>0$. Such
exotic matter can be modelled by a scalar field having a positive
energy (dark energy) or negative energy (phantom) (Caldwell
\emph{et al.} ). Phantom scalar field can be motivated by S-brane
arising in string theory (Townsend and Wohlfarth). Phantom fields
are introduced by bulk viscosity ($\eta$) effects that are
equivalent to replacing the pressure $p$ bye an effective
pressure, viz., $p_{\rm eff.}=p-3\eta H$, where $H$ is the Hubble
constant. Viscous effects in an expanding universe are connected
with dissipations that are attributed to creation of energy
(matter) in the universe.

Several authors  suggested that the bulk viscosity can drive the
universe into a period of exponential expansion (inflation)
(Murphy, Gr$\rm\phi$n, Beesham, Arbab). This is really the case,
as the effect of bulk viscosity in an expanding universe is to
decrease the pressure making the total pressure negative.
Inflation can also be induced by higher order corrections
(Starobinsky, 1980). In scalar field, inflation stopped by the
slow roll-down of the scalar field from the false vacuum to the
true vacuum. In this work, we investigate the coupling of gravity
with vacuum and viscosity. Provided a certain conspiracy is
maintained the evolution of the universe can proceed in an
attractive way. We have found that such a recipe is possible and
leads to interesting features related to the cosmic evolution.
Inflation is triggered by the vacuum energy bulk viscosity
cooperation. This work generalizes our recent model of phantom and
dark energy models (Arbab, 2007 ).

\section{The Model}
Consider the Einstein-Hilbert action with a cosmological constant
($\Lambda$)
\begin{equation}
S=-\frac{1}{16\pi G}\int d^4x\sqrt{g}(R+2\Lambda)+S_{\rm matter}
\end{equation}
The variation of the metric with respect to $g_{\mu\nu}$ with
$f(R)=R-2\Lambda$,  gives (Amarzguioui \emph{et al.})
\begin{equation}
f'(R)R_{\mu\nu}-\frac{1}{2}f(R)g_{\mu\nu}=-8\pi G T_{\mu\nu}\ ,
\end{equation}
where $T_{\mu\nu}$ is the energy momentum tensor of the cosmic
fluid.
\\
For an ideal fluid one has
\begin{equation}
T_{\mu\nu}=(\rho+p)u_\mu u_\nu+pg_{\mu\nu},
\end{equation}
where $u_\mu, \rho, p$ are the velocity, density and pressure of
the cosmic fluid. \\
Contracting Eq.(2), using Eq.(3) and taking its 00 components give
the equation,
\begin{equation}
 Rf'(R)-2f(R)+8\pi GT=0,
\end{equation}
and
\begin{equation}\label{2}
f'(R)R_{00}+\frac{1}{2}f(R)+8\pi G\ T_{00}=0,
\end{equation}
with $T_{00}=\rho$, $T=\rho-3p$ and $T_{ij}=-p$ for $i,j=1,2,3$.\\
For a flat Friedmann-Lemaître-Robertson-Walker metric,
$$ds^2=dt^2-a^2(t)\left(dr^2+r^2(d\theta^2+\sin^2\theta\ d\phi^2)\right),$$
one has $R_{00}=-3\left(\frac{\ddot a}{a}\right)$ and
$R=-6\left((\frac{\dot a}{a})^2+\frac{\ddot a}{a}\right)$, where
$a$ is the scale factor, so that Eqs.(4) and (5) yield
\begin{equation}\label{1}
 3\left(  \frac{\dot a}{a}\right)^2=8\pi \ G\rho +\Lambda,
\end{equation}
\begin{equation}\label{1}
3 \left(  \frac{\ddot a}{a}\right)=-4\pi\ G(\rho+3p) +\Lambda,
\end{equation}
and the energy conservation equation reads,
\begin{equation}\label{1}
   \dot\rho+3\left(\frac{\dot a}{a}\right)(\rho+p)=0.
\end{equation}
The pressure $p$ and energy density $\rho$ of an ideal fluid are
related by the equation of state,
\begin{equation}
p=\omega\ \rho,\qquad\omega=\rm const.
\end{equation}
The Einstein field equation, with time-dependent $G$ and
$\Lambda$, then yields two independent equations [Eqs.(6) and (7)]
having the same form as in the standard model. Hence, we can now
allow $\Lambda$ and $G$ to vary with time, i.e.,
$\Lambda=\Lambda(t)$ and $G=G(t)$.

The Bianchi identity
\begin{equation}\label{1}
(R^{\mu\nu}-\frac{1}{2}Rg^{\mu\nu})_{;\ \mu}=-(8\pi
GT^{\mu\nu}+\Lambda g^{\mu\nu})_{;\ \mu}=0,
\end{equation}
with Eqs.(2) and (3) imply that
\begin{equation}\label{1}
 G\dot\rho+3(p+\rho) G\frac{\dot a}{a}+\rho\dot
G+\frac{\dot\Lambda}{8\pi}=0\ ,
\end{equation}
Bulk viscosity can be introduced in a uniform perfect fluid by
replacing the pressure term $p$ by an effective pressure, $p_{\rm
eff.}$ defined by
\begin{equation}
p_{\rm eff.}=p-3\eta H\ ,
\end{equation}
where,  $\eta$ is the coefficient of bulk viscosity. This is
normally modelled by the relation
\begin{equation}
\eta=\eta_0\rho^n\ ,\qquad n\ , \ \eta_0 =\rm const.
\end{equation}
Applying Eq.(12) into Eq(11) and using Eq.(8), one obtains (Arbab,
1997)
\begin{equation}\label{1}
8\pi \dot G\rho+\dot\Lambda=9\eta (8\pi G)H^2.
\end{equation}
We consider here the ansatz
\begin{equation}\label{1}
   \Lambda=\frac{3\beta }{\rho^\alpha} \ ,\qquad \beta\ , \ \alpha=\rm const.
\end{equation}
Integrating Eq.(11), using Eq.(12), we obtain
\begin{equation}\label{0}
    \rho=A a^{-3(1+\omega)}\ , \ A=\rm const.
\end{equation}
Using Eqs.(6) and (15), Eq.(14) reads
\begin{equation}\label{1}
 \frac{\dot G}{G}-\frac{3\alpha\beta}{8\pi G \rho^{(\alpha+1)}}\left(\frac{\dot\rho}{\rho}\right)=3\eta_0\rho^n(8\pi
 G\rho+3\beta\rho^{-\alpha}),
\end{equation}
Now  consider the following functional dependence of the
gravitational constant:
\begin{equation}\label{0}
  8\pi G=C\rho^{-(\alpha+1)} \ , \ C=\rm const.
\end{equation}
Eqs.(17) and (18) imply that
\begin{equation}\label{0}
  \dot\rho=N\rho^{n-\alpha+1},\qquad
  N=-\frac{3C\eta_0(C+3\beta)}{C(1+\alpha)+3\alpha \beta}.
\end{equation}
so that
\begin{equation}\label{0}
  \frac{\dot a}{a}=K\
  a^{-3(1+\omega)(n-\alpha)},
\end{equation}
where $K=-\frac{N}{3}\frac{A^{(n-\alpha)}}{(1+\omega)}\ , \ \
\omega\ne -1$. Substituting this in Eq.(13) using Eq.(14), one
gets
\begin{equation}
 a=D \ t^{1/3(1+\omega)(n-\alpha)},
 \end{equation}
 where $D=A^{1/3(1+\omega)}\left[-N(n-\alpha)\right]^{1/3(1+\omega)(n-\alpha)},
 \ \ n\ne\alpha.$
Substituting the above equation in Eq.(16), one finds
\begin{equation}\label{5}
 \rho=\left[-N(n-\alpha)\right]^{-1/(n-\alpha)}\
 t^{-1/(n-\alpha)},\qquad n\ne\alpha
\end{equation}
so the Eq.(15) beomes
\begin{equation}\label{5}
\Lambda=3\beta\left[-N(n-\alpha)\right]^{\alpha/(n-\alpha)}\
t^{\alpha/(n-\alpha)},\qquad n\ne\alpha.
\end{equation}
Eq.(18) now reads
\begin{equation}\label{5}
G=\frac{C}{8\pi}\left[-N(n-\alpha)\right]^{(1+\alpha)/(n-\alpha)}\
t^{(1+\alpha)/(n-\alpha)},\qquad n\ne\alpha.
\end{equation}
\section{Phantom energy}
Consider now the following cases:
\subsection{Case (1)}
Now let $n=\frac{\alpha}{2}$ where $-1<\alpha<0$ and $1+\omega>0$.
In this case, Eqs.(21), (22),  reduce to
\begin{equation}\label{5}
a\propto t^{-2/3\alpha(1+\omega)},\qquad \rho\propto t^{2/\alpha},
\end{equation}
and Eqs. (23), (24)  and (13) yield
\begin{equation}
 G\propto t^{-2(1+\alpha)/\alpha}, \qquad
\Lambda\propto t^{-2}, \qquad \eta\propto t,
\end{equation}
where $C>3\beta>0$, i.e., $G>0$. These represent the viscous
analogue of the dark energy model (Arbab, 2007). In particular, a
viscous cosmological model with $\Lambda\propto H^2$ (Arbab, 1997)
is equivalent to a viscous dark  energy model with $\Lambda\propto
\rho^{-\alpha}$ if $n_{\rm arb.}=1+\frac{\alpha}{2}$., where
$n_{\rm arb.}$ is the index of viscosity in (Arbab, 1997).
\subsection{Case (2)}
Now let  $\alpha>0$ and $n<\alpha$. In this case Eq.(19) implies
that one has the phantom energy solution, viz., $1+\omega<0$. One
requires here $C<0$ so that $G<0$, and for $\beta>0$ one has
$\Lambda>0$. This solution is found by (Arbab, 2007) and the above
solution represents its viscous analogue. It is clear here that
though the energy density increases, gravity (decreases) and
viscosity (increase) conspire not to allow the phantom energy
density to dominate.
\section{Inflationary Solution}
We notice from Eq.(20) that when $n=\alpha>0$, we obtain
\begin{equation}
\frac{\dot a}{a}=K=\frac{C\eta_0(C+3\beta)}{C(1+\alpha)+3\alpha
\beta}\frac{1}{(1+\omega)}.
\end{equation}
 This implies that
 \begin{equation}
 a=\Gamma \exp(Kt)\ , \qquad \Gamma={\rm const.},
\end{equation}
where $K>0$, for $1+\omega>0$.
 Eqs.(16),  yield
\begin{equation}\label{5}
\rho\propto \exp[-3K(1+\omega)t]\ ,
\end{equation}
so that Eqs. (15) and (18)   become
\begin{equation}
\Lambda\propto\exp[3\alpha K(1+\omega)t]\ \ ,\ \
G\propto\exp[3K(1+\omega)(1+\alpha)t]\ ,
\end{equation}
and the bulk viscosity, Eq.(13),
\begin{equation}
\eta\propto\exp[-3\alpha K(1+\omega)t] .
\end{equation}
Notice, however, during inflation $\omega\ne -1$ as evident from
Eq.(24). This is unlike the standard case where inflation requires
$\omega=-1$. We need a 60 e-folding for successful inflation. It
is also remarkable that the cosmological constant during inflation
increases exponentially with time, whereas the energy density
decreases exponentially with time. Whatever the initial value of
the cosmological constant before inflation, its value at the end
of inflation will be enormously large.   This may explain why
$\Lambda$ was large in the very early universe, as suggested from
particle physics considerations. The universe exited from
inflation due to the huge growth of the gravitational force that
halt the exponential expansion. Thereafter, the universe enters a
radiation dominated phase. The large decrease of the bulk
viscosity during the inflationary era has allowed the universe to
isotropize, and eventually led to the isotropic and homogenous
universe, we observe today. In the standard de-Sitter model the
inflationary expansion is led by the cosmological constant
($\Lambda)$, where the energy density stays constant. We notice
from Eqs. (28) and (29) that
 the mass created (annihilated) during inflation is $M\propto \rho
 a^3\propto\exp(-3K\omega t)$. But since $\omega>-1$, one has for
 $-1<\omega<0$, a positive mass creation rate. Hence, one would presume
 that all matter constituting the universe mass was produced during inflation decelerated inflation.

 It is remarkable to notice that inflation is induced by dark
 energy only. Thus, dark energy played an important role  by driving
 the early universe into an exponential expansion, and the present universe
 into a accelerated expansion. Hence, the existence of
 dark energy is very crucial to the evolution of the universe.
\section{Concluding Remarks}
We have studied in this work the effect of bulk viscosity on the
evolution of dark matter and phantom energies. We have shown that
non-viscous dark matter models are equivalent to viscous ones. The
increasing bulk viscosity and decreasing gravitational constant do
not allow the phantom energy density to condensate. During
inflationary era the universe isotropizes and the cosmological
constant  attained a vary large value. After inflation the
cosmological constant decreases with time quadratically (e.g., for
$n=\frac{\alpha}{2}$). This evolution  provides a viable mechanism
for the smallness of the present cosmological constant, i.e.,  why
today the cosmological constant is vanishingly small compared with
its inial value!
\section{Acknowledgments}
I am grateful to the Swedish International Development Agency
(SIDA) for providing financial support for my visit to the Abdus
Salam International Center for Theoretical Physics, where this
work is carried out.
\section{References}
 A. Beesham, \emph{Int. J. Theor. Phys.}, 25, 1295 (1986).\\
R. Caldwell, M. Kamionkowski and N. Weinberg, \emph{Phys. Rev.
Lett}. 91, 071301 (2003). \\
R. Caldwell, \emph{Phys.Lett.B}545, 23 (2002).\\
S. Perlmutter, \emph{et al.}, \emph{Ap. J}, 517, 565 (1998).\\
A. Reiss, \emph{et al.}, \emph{Ap. J}, 116, 1009 (1998).\\
A. I. Arbab, \emph{Gen. Relt. Gravit.}, 29, 61 (1997).\\
P.K. Townsend and M.N.R. Wohlfarth, \emph{Phys. Rev. Lett.} 91,
061302 (2003).\\
G.L. Murphy, \emph{Phys. Rev.} D 8, 4231 (1988)\\
$\rm \phi$. Gr$\rm \phi$n,  \emph{Astrophys. \& Space Sci.}173, 191 (1990).\\
A.A. Starobinsky, \emph{Phys. Lett.} 91B,   99 (1980).\\
A. I. Arbab, \emph{hep-th/0711.1465v1} (2007).
\end{document}